\documentclass[a4paper,10pt]{article}
\usepackage[utf8x]{inputenc}
\usepackage{amssymb}
\usepackage{graphicx}
\usepackage{dcolumn}
\usepackage{bm}
\usepackage{amsmath}
\title{Robust Short-Term Memory without Synaptic Learning}
\author{Samuel Johnson$^{1,2,}$\thanks{samuel.johnson@imperial.ac.uk}, J. Marro$^3$, and Joaqu\'{\i}n J. Torres$^3$
\\
\small{$^1$Department of Mathematics, Imperial College London, SW7 2AZ, United Kingdom.} 
\\
\small{$^2$Oxford Centre for Integrative Systems Biology, and Department of Physics,}
\\
\small{University of Oxford, Clarendon Lab, OX1 3QU, United Kingdom.}
\\
\small{$^3$Departamento de Electromagnetismo y F\'{\i}sica de la Materia, and}
\\
\small{Institute
\textit{Carlos I} for Theoretical and Computational Physics,}\\
\small{University of Granada, 18071 Granada, Spain.}
}

\date{}

\begin{document}

\maketitle

\begin{abstract}
Short-term memory in the brain cannot in general be explained the way long-term memory can -- as a gradual modification of synaptic weights -- since it takes place too quickly. Theories based on some form of cellular bistability, however, do not seem able to account for the fact that noisy neurons can collectively store information in a robust manner. We show how a sufficiently clustered network of simple model neurons can be instantly induced into metastable states capable of retaining information for a short time (a few seconds). The mechanism is robust to different network topologies and kinds of neural model. This could constitute a viable means available to the brain for sensory and/or short-term memory with no need of synaptic learning. Relevant phenomena described by neurobiology and psychology, such as local synchronization of synaptic inputs and power-law statistics of forgetting avalanches, emerge naturally from this mechanism, and we suggest possible experiments to test its viability in more biological settings.
\end{abstract}


%Johnson S, Marro J, Torres JJ (2013) Robust Short-Term Memory without Synaptic Learning. PLoS ONE 8(1): e50276. doi:10.1371/journal.pone.0050276
%http://www.plosone.org/article/info:doi/10.1371/journal.pone.0050276

%nonequilibrium neural networks.

\section*{Author summary}

Whenever an image is flashed briefly before your eyes, or you hear a sudden sound, you are usually able to recall the information presented for a few seconds thereafter. In fact, it is most vivid at first but fades gradually. According to our current understanding of neural networks, memories are stored by strengthening and weakening the appropriate connections (synapses) between neurons. But these biochemical processes take place on a timescale of minutes. Therefore, when it came to understanding such short-term memory tasks, it seemed that either each neuron had an individual memory that could work fast enough (not very robust), or we could only actually remember things that had been stored in our brains previously (not very credible). Here we suggest a simple mechanism -- Cluster Reverberation -- whereby neurons with no individual memory can nonetheless store completely novel information almost instantly and maintain it for a few seconds, thanks to a preponderance of local connections in the network. If the brain were indeed using this mechanism, it might explain the statistics of forgetting as well as some recent neurobiological findings.

\section*{Introduction}
\label{sec_intro}
\subsection*{Slow but sure, or fast and fleeting?}

%``As long as the brain is a mystery, the universe will remain a mystery'' -- or so, at least, contented Santiago Ram\'on y Cajal. We have come some way towards unraveling this mystery since he showed that the brain is basically a network of neurons. For instance, we now have an idea

Memory
 -- the storage and retrieval of information by the brain --
is probably nowadays one of the best understood of all the collective phenomena to emerge in that most complex of systems.
Thanks to a gradual modification of synaptic weights (the interaction strengths with which neurons signal to one other) particular patterns of firing and non-firing cells become energetically favourable and so systems evolve towards these attractors according to a mechanism known as Associative Memory \cite{Hebb,Amari,Hopfield,Amit}. In nature, these synaptic modifications occur via the biochemical processes of long-term potentiation (LTP) and depression (LTD) \cite{Gruart,DeRoo}. However, some memory processes take place on timescales of seconds or less
and in many instances cannot be accounted for by LTP and LTD \cite{Durstewitz}, since these require at least minutes to be effected \cite{Lee,Klintsova}.
For example,
visual stimuli are recalled in great detail for up to about one second after exposure (iconic memory); similarly, acoustic information seems to linger for three or four seconds (echoic memory) \cite{Sperling,Cowan}. 
In fact, it appears that the brain actually holds and continually updates a kind of buffer in which sensory information regarding its surroundings is maintained (sensory memory) \cite{Baddeley_Book}. This is easily observed by simply closing one's eyes and recalling what was last seen, or thinking about a sound after it has finished.
Another instance is the capability referred to as \textit{working} memory \cite{Durstewitz, Baddeley}: just as a computer requires RAM for its calculations despite having a hard drive for long-term storage, the brain must continually store and delete information to perform almost any cognitive task. We shall 
here use {\it short-term} memory
to describe the brain's ability to store information on a timescale of seconds or less.
%\footnote{This is not a standard definition. In fact, sensory memory is usually considered distinct from short-term memory, but we shall use the latter generically since the mechanism we propose in this paper could be relevant for either or both phenomena. We should also point out that the recent flurry of research in psychology and neuroscience on working memory has lead to this term sometimes being used to mean short-term memory; strictly speaking, however, working memory is generally considered to be an aspect of cognition which operates on information stored in short-term memory.}

Evidence that short-term memory is related to sensory information while long-term memory is more conceptual can be found in psychology. For instance, a sequence of similar sounding letters is more difficult to retain for a short time than one of phonetically distinct ones, while this has no bearing on long-term memory, for which semantics seems to play the main role \cite{Conrad_1,Conrad_2}; and the way many of us think about certain concepts, such as chess, geometry or music, is apparently quite sensorial: we imagine positions, surfaces or notes as they would look or sound.
Most theories of short-term memory -- which almost always focus on working memory -- make use of some form of previously stored information (i.e., of synaptic learning) and so can account for
labelling tasks, such as remembering a particular series of digits or a known
% Revised
word,\footnote{This method can also be used to represent a continuous variable, such as the value of an angle or the length of an object, because concepts such as {\it angle} and {\it length} are in some sense already ``known'' at the time of the stimulus \cite{Wang}.} 
but not for the instant recall of novel
information
\cite{Barak,Roudi,Mongillo}.
An interesting exception is the mechanism proposed by Chialvo {\it et al.} \cite{Chialvo_refractory} which allows for arbitrary patterns of activity to be temporarily retained thanks to the refractory times of neurons.

Attempts to deal with novel information
%the latter 
% end revision
have been made by proposing mechanisms of \textit{cellular bistability}: neurons are assumed to retain the state they are placed in (such as firing or not firing) for some period of time thereafter \cite{Camperi,Fukai,Tarnow}.
Although there may indeed be subcellular processes leading to a certain bistability, the main problem with short-term memory depending exclusively on such a mechanism is that if each neuron must act independently of the rest the patterns will not be robust to random fluctuations \cite{Durstewitz} -- and the behaviour of individual neurons is known to be quite noisy \cite{Compte}. It is worth pointing out that one of the strengths of Associative Memory is that the behaviour of a given neuron depends on many neighbours and not just on itself, which means that robust global recall can emerge despite random fluctuations at an individual level. 

\subsection*{Harnessing network structure}

Something that, at least until recently, most neural-network models have failed to take into account is the structure of the network -- its topology -- it often being assumed that synapses are placed  among the neurons completely at random, or even that all neurons are connected to all the rest.
Although relatively little is yet known about the architecture of the brain at the level of neurons and synapses, experiments have shown that it is heterogeneous (some neurons have very many more synapses than others), clustered (two neurons have a higher chance of being connected if they share neighbours than if not) and highly modular (there are groups, or modules, with neurons forming synapses preferentially to those in the same module) \cite{Sporns, Johnson_JSTAT}.
We show here that it suffices to use a more realistic network topology, in particular one that is modular and/or clustered, for a randomly chosen pattern of activity the system is placed in to be metastable. This means that novel information can be instantly stored and retained for a short period of time in the absence of
both
synaptic learning
and
cellular bistability. The only requisite is that the patterns be coarse grained versions of the usual patterns -- that is, whereas it is
often assumed that each neuron 
in some way
represents one bit of information, we shall allocate a bit to a small group or neurons. (This does not, of course, mean that memories are expected to be encoded as bitmaps. In fact, we are not making any assumptions regarding neural coding.)
%\footnote{This does not, of course, mean that memories are expected to be encoded as bitmaps. Just as with individual neurons, positions or orientations, say, could be represented by the activation of particular sets of clusters. In fact, we are not making any assumptions regarding neural coding.}

The mechanism, which we call Cluster Reverberation (CR), is very simple. If neurons in a group are more densely connected to each other than to the rest of the network, either because they form a module or because the network is significantly clustered, they will tend to retain the activity of the group: when they are all initially firing, they each continue to receive many action potentials and so go on firing, whereas if they start off silent, there is not usually enough input current from the outside to set them off.
% Revised:
(This is similar to the `re-entrant' activity exhibited by excitable elements \cite{Lewis}.)
% end revision
The fact that each neuron's state depends on its neighbours confers to the mechanism a certain robustness to random fluctuations. This robustness is particularly important for biological neurons, which as mentioned are quite noisy. Furthermore, not only does the limited duration of short-term memory states emerge naturally from this mechanism (even in the absence of interference from new stimuli) but this natural forgetting follows power-law statistics, as has been observed experimentally \cite{Wixted_1, Wixted_2, Sikstrom}.
It is also coherent with recent observations of locally synchronized neural activity {\it in vivo} \cite{Local_sync}, and of clustering in both synaptic inputs \cite{Kleindienst} and plasticity \cite{Makino} during development. The viability of this mechanism in a more realistic setting could perhaps be put to the test by growing modular and/or clustered networks {\it in vitro} and carrying out similar experiments as we do here in simulation \cite{Ole2,Shein} (see Discussion).

%The process is reminiscent both of block attractors in ordinary neural networks \cite{Dominguez} and of domains in magnetic materials \cite{Hubert}, while Mu\~noz et al. have recently highlighted a similarity with Griffiths phases on networks \cite{Munoz}. It can also be interpreted as a multiscale phenomenon: the mesoscopic clusters take on the role usually played by individual neurons, yet make use of network properties.

\section*{Results}
\label{sec_model}
\subsection*{The simplest neurons on modular networks}

Consider a network of $N$
model neurons, with activities $s_{i}=\pm 1$.
The topology is given by the adjacency matrix $\hat{a}_{ij}=\lbrace 1,0\rbrace$, each element representing the existence or absence of a synapse from neuron $j$ to neuron $i$ ($\hat{a}$ need not be symmetric). In this kind of model --  a network of what are often referred to as Amari-Hopfield neurons -- each edge usually has a \textit{synaptic weight} associated, $\omega_{ij}\in\mathbb{R}$, which serves to store information \cite{Hebb,Amari,Hopfield,Amit}. However, since our objective is to show how this can be achieved without synaptic learning, we shall here consider these to have all the same value: 
% Revised: omega>0
$\omega_{ij}=\omega>0$ $\forall i,j$. Neurons are updated in parallel (Little dynamics) at each time step, according to the stochastic transition rule
\begin{equation}
P(s_{i}\rightarrow \pm1)= \frac{1}{2} \left[\pm\tanh\left(\frac{h_{i}}{T}\right)+1\right],
\label{eq_P}
\end{equation}
where
$h_i=\omega\sum_j{\hat{a}}_{ij}s_{j}$
is the {\it field} at neuron $i$,
and $T$ is a stochasticity parameter called \textit{temperature}.
This dynamics can be derived by considering coupled binary elements in a thermal bath, the transition rule stemming from energy differences between states \cite{Hopfield,Amit,Fraiman}.

We shall consider the network defined by $\hat{a}$ to be made up of $M$ distinct modules. To achieve this, we can first construct $M$ separate random directed networks, each with $n=N/M$ nodes and mean degree (mean number of neighbours) $\langle k\rangle$. Then we evaluate each edge
$\hat{a}_{ij}=1$ and, with probability $\lambda$, eliminate it ($\hat{a}_{ij}\rightarrow0$), to be substituted for another edge between the original (postsynaptic) neuron $i$ and a new (presynaptic) neuron $l$ chosen at random from among any of those in other modules
%\footnote{We do not allow self-edges (although they can occur in reality) since these could be regarded as equivalent to a form of cellular bistability.}
($\hat{a}_{il}\rightarrow1$). We do not allow self-edges (although they can occur in reality) since these could be regarded as equivalent to a form of cellular bistability.
Note that this protocol does not alter the number of presynaptic neighbours of each node, $k^{in}_{i}=\sum_{j}\hat{a}_{ij}$, although the number of postsynaptic neurons, $k^{out}_{i}=\sum_{j}\hat{a}_{ji}$, can vary.
The parameter $\lambda$ can be seen as a measure of  {\it modularity} of the partition considered, since it coincides with the expected value of the proportion of edges that link different modules \cite{Newman_rev}. In particular, $\lambda=0$ defines a network of disconnected modules, while $\lambda=1-M^{-1}$ yields a random network in which this partition has no modularity. If $\lambda\in(1-M^{-1},1)$, the partition is less than randomly modular -- i.e., it is \textit{quasi-multipartite} (or multipartite if $\lambda=1$).

\subsection*{Cluster Reverberation}
\label{sec_Cluster_Reverb}

\begin{figure}[htb!]
\begin{center}
\includegraphics[scale=0.4]{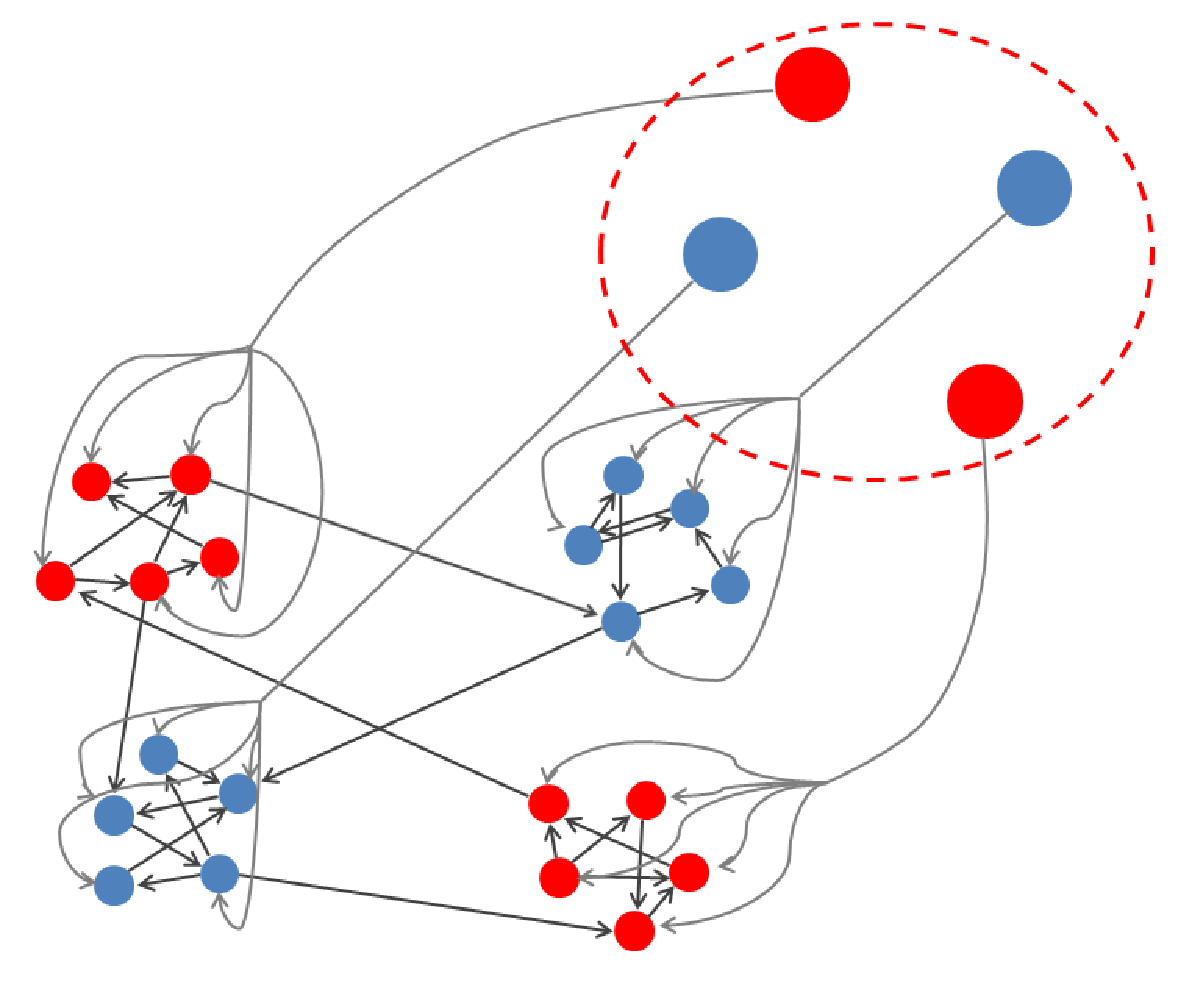}
\end{center}
\caption{Diagram of a modular network composed of four five-neuron clusters. The four circles enclosed by the dashed line represent the stimulus: each is connected to a particular module, which adopts the input state (red or blue) and retains it after the stimulus has disappeared thanks to Cluster Reverberation.}
\label{fig_picture}
\end{figure}

A memory pattern, in the form of a given configuration of activities, $\lbrace \xi_{i} =\pm1\rbrace$, can be stored in this system with no need of prior learning.
(The system will recall the pattern perfectly when $s_i=\xi_i$, $\forall i$.)
Imagine a pattern
such that the activities of all $n$ neurons found in any module are the same -- i.e., $\xi_{i}=\xi_{\mu(i)}$, where the index $\mu(i)$ denotes the module that neuron $i$ belongs to. 
The system can be induced into this configuration through the application of an appropriate stimulus: the field of each neuron will be altered for just one time step according to
$$%\begin{equation}
h_{i}\rightarrow h_{i}+\delta \xi_{\mu(i)}, \forall i,
%\label{h_d}
$$ %\end{equation}
where the factor $\delta$ is the intensity of the stimulus (see Fig. \ref{fig_picture}). This mechanism for dynamically storing information will work for values of parameters such that the system is sensitive to the stimulus, acquiring the desired configuration, yet also able to retain it for some interval of time thereafter (a similar setting is considered, for instance, in Ref. \cite{Johnson_EPL}).

% Revised:
%The two main attractors of the system are $s_{i}=1$ $\forall i$ and $s_{i}=-1$ $\forall i$. These are the configurations of minimum energy
The two configurations of minimum energy of the system are $s_{i}=1$ $\forall i$ and $s_{i}=-1$ $\forall i$ (see the next section for a more detailed discussion on energy).
However, the energy is locally minimized for any configuration in which
each module comprises either all active or all inactive neurons (that is, for configurations $s_{i}=d_{\mu(i)}$ $\forall i$, with $d_{\mu(i)}=\pm 1$
a binary variable specific to the whole module $\mu(i)$ that neuron $i$ belongs to).
%which $s_{i}=d_{\mu(i)}$ $\forall i$ with $d_{\mu}=\pm 1$; that is, configurations such that each module comprises either all active or all inactive neurons.
% End revision.
These are the configurations that we shall use to store information. We define the mean activity
%\footnote{The mean activity in a neural network model is usually taken to represent the mean firing rate measured in experiments \cite{Torres_rev}.}
of each module, $m_{\mu}\equiv \langle s_{i}\rangle_{i\in \mu}$, 
which is a mesoscopic variable, as well as the global mean activity,
$m\equiv\langle s_{i}\rangle_{\forall i}$
(these magnitudes change with time, but, where possible, we shall avoid writing the time dependence explicitly for clarity; $\langle \cdot\rangle_{x}$ stands for an average over $x$).
The mean activity in a neural network model is usually taken to represent the mean firing rate measured in experiments \cite{Torres_rev}.
The extent to which the network, at a given time, retains the pattern $\lbrace\xi_{i}\rbrace$ with which it was stimulated is measured with the \textit{overlap} parameter
$m_{stim}\equiv\langle\xi_{i}s_{i}\rangle_{i}=\langle\xi_{\mu}m_{\mu}\rangle_{\mu}$.
Ideally, the system should be capable of reacting immediately to a stimulus by adopting the right configuration, yet also be able to retain it for long enough to use the information once the stimulus has disappeared. A measure of performance for such a task is therefore
$$%\begin{equation}
\eta\equiv\frac{1}{\tau}\sum_{t=t_{0}+1}^{t_{0}+\tau}m_{stim}(t),
$$%\end{equation}
where $t_{0}$ is the time at which the stimulus is received and $\tau$ is the period of time we are interested in ($|\eta|\leq1$) \cite{Johnson_EPL}. If the intensity of the stimulus, $\delta$, is very large, then the system will always adopt the right pattern perfectly and $\eta$ will only depend on how well it can then retain it. In this case, the best network will be one
that is
made up of
mutually disconnected modules ($\lambda=0$). However, since the stimulus in a real brain can be expected to arrive via a relatively small number of axons, either from another part of the brain or directly from sensory cells, it might be more realistic to assume that $\delta$ is of a similar order as the input a typical neuron receives from its neighbours, $\langle h\rangle \sim\omega \langle k\rangle$.

\begin{figure}[htb!]
\begin{center}
\includegraphics[scale=0.5]{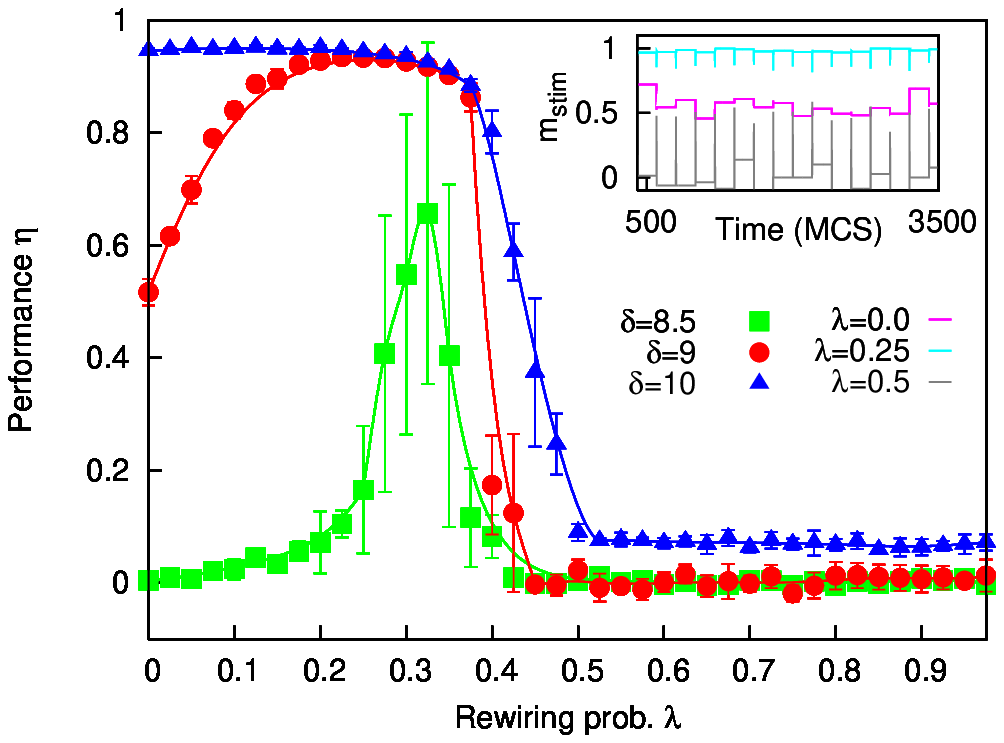}%{fig1.eps}
\end{center}
\caption{
Performance $\eta$ against $\lambda$ for networks of the sort described in the main text with $M=160$ modules of $n=10$ neurons each, $\langle k\rangle=9$, obtained from Monte Carlo (MC) simulations; patterns are shown with intensities $\delta=8.5$, $9$ and $10$, 
% Revised:
and performance is computed evey $200$ time steps, preceding the next random stimulus; 
$T=0.02$
(error bars represent standard deviations; lines -- splines -- are drawn as a guide to the eye). Inset: typical time series of $m_{stim}$ (i.e., the overlap with whichever pattern was last shown) for $\lambda=0.5$ (bad performance), $0$ (intermediate), and $0.25$ (optimal); with $\delta=\omega \langle k\rangle=9$.}
\label{fig_performance}
\end{figure}

Figure \ref{fig_performance} shows the mean performance obtained in Monte Carlo (MC) simulations when the network is repeatedly stimulated with different randomly generated patterns. For low enough values of $\lambda$ and stimuli of intensity $\delta \gtrsim \omega\langle k\rangle$, the system can capture and successfully retain any pattern it is ``shown'' for some period of time, even though this pattern was in no way previously learned. For less intense stimuli ($\delta<\omega\langle k\rangle$), performance is nonmonotonic with modularity: there exists an optimal value of $\lambda$ at which the system is sensitive to stimuli yet still able to retain new patterns quite well.

% Revised:
Just as some degree of structural (quenched) noise, given by $\lambda$, can improve performance by increasing sensitivity, so too the dynamical (annealed) noise set by $T$ can have a similar effect. This apparent stochastic resonance is analysed in Analysis: The effect of noise.
%the Appendix.
% end revision

\subsection*{Energy and topology}
\label{sec_energy}

\begin{figure}[htb!]
\begin{center}
\includegraphics[scale=0.5]{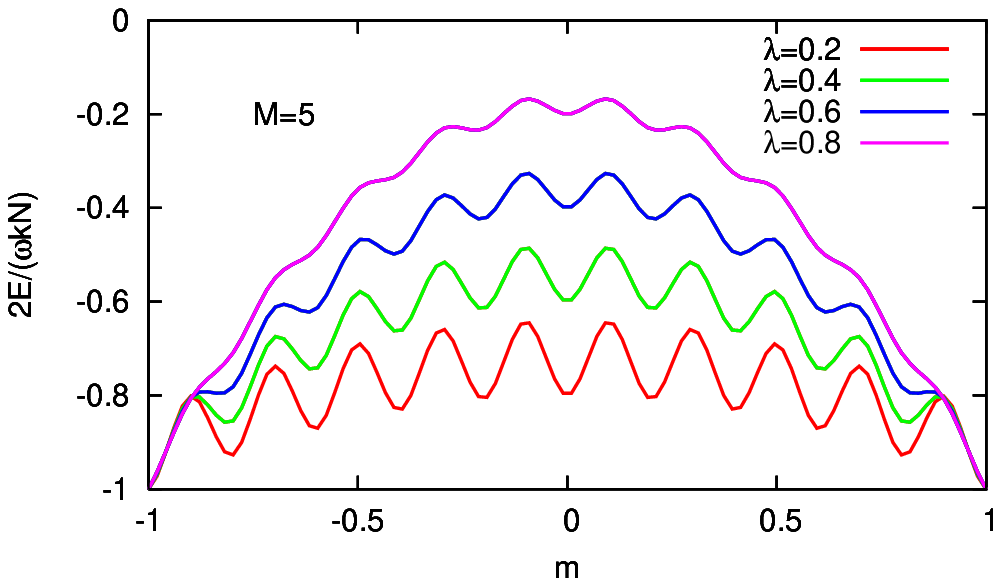}%{fig2.eps}
\end{center}
\caption{
%(colour online) 
Configurational energy of a network made up of $M=5$ modules of $n=10$ neurons each, according to Eq. (\ref{E}), for various values of $\lambda$ (increasing from bottom to top). The minima correspond to situations such that all neurons within any given module have the same sign.
}
\label{fig_energy}
\end{figure}

Each pair of neurons contributes a configurational energy $e_{ij}=-\frac{1}{2}\omega (\hat{a}_{ij}+\hat{a}_{ji})s_{i}s_{j}$ \cite{Amit}; that is, if there is an edge from $i$ to $j$ and  they have opposite activities, the energy is increased in $\frac{1}{2}\omega$, whereas it is decreased by the same amount if their activities are equal. Given a configuration, we can obtain its associated energy  by summing over all pairs.
To study how the system relaxes from the metastable states (i.e., how it ``forgets'' the information stored)
we shall be interested in configurations with $x$ neurons that have $s=+1$ (and $N-x$ neurons with $s=-1$), chosen in such a way that one module at most, say $\mu$, has neurons in both states simultaneously. Therefore, $x=n\rho+z$, where $\rho$ is the number of modules with all their neurons in the positive state and $z$ is the number of neurons with positive sign in module $\mu$. We can write $m=(2x-1)/N$ and $m_{\mu}=(2z-1)/n$. The total configurational energy of the system will be 
$$
E=\sum_{ij}e_{ij}=\frac{1}{2}\omega(L_{\uparrow\downarrow}-\langle k\rangle N),
$$
where $L_{\uparrow\downarrow}$ is the number of edges linking nodes with opposite activities. By simply counting over expected numbers of edges, we can obtain the expected value of $L_{\uparrow\downarrow}$ (which amounts to a mean-field approximation),
yielding:
\begin{eqnarray}
\frac{E}{\omega\langle k\rangle}+\frac{N}{2}= 
\frac{\lambda n}{N-n}\lbrace \rho[n-z+n(M-\rho-1)]
\label{E}
\\
+(M-\rho-1)(z+n\rho)\rbrace
%-\frac{1}{2}N
+(1-\lambda)\frac{z(n-z)}{n-1}
\nonumber
.
%\label{E}
\end{eqnarray}
Figure \ref{fig_energy} shows the mean-field configurational energy curves for various values of the modularity on a small modular network. The local minima (metastable states) are the configurations used to store patterns. It should be noted that the mapping $x\rightarrow m$ is highly degenerate: there are $C_{mM}^{M}$ patterns with mean activity $m$ that all have the same energy.

\subsection*{Forgetting avalanches}
\label{sec_avalanches}

In obtaining the energy we have assumed that the number of synapses rewired from a given module is always equal to its expected value: $\nu=\langle k\rangle n\lambda$. However, since each edge is evaluated with probability $\lambda$, $\nu$ will in fact vary somewhat from one module to another, being approximately Poisson distributed with mean $\langle \nu\rangle=\langle k\rangle n\lambda$. The depth of the energy well corresponding to a given module is then, neglecting all but the last term in Eq. (\ref{E}) and approximating $n-1\simeq n$, $\Delta E\simeq \frac{1}{4}\omega(n\langle k\rangle -\nu)$.
The typical escape time $\tau$ from an energy well of depth $\Delta E$ at temperature $T$ is $\tau\sim e^{\Delta E/T}$ \cite{Levine}. Using Stirling's approximation 
[$n!\sim \sqrt{2\pi n}(n/e)^n$]
in the Poisson distribution over $\nu$ and expressing it in terms of $\tau$, we find that the escape times are distributed according to
\begin{equation}
P(\tau)\sim \left(1 -\frac{4T}{\omega n\langle k\rangle}
\ln\tau\right)^{-\frac{3}{2}}\tau^{-\beta(\tau)},
\label{eq_Pt}
\end{equation}
where
\begin{equation}
\beta(\tau)=1+\frac{4T}{\omega n\langle k\rangle}\left[1+\ln\left(\frac{\lambda n\langle k\rangle}{1-\frac{4T}{\omega n\langle k\rangle}\ln\tau}  \right) \right].
\label{eq_beta}
\end{equation}
Therefore, at low temperatures, $P(\tau)$ will behave approximately like a power law.
% Revised:
Note also that the size of the network, $N$, does not appear in Eqs. (\ref{eq_Pt}) and (\ref{eq_beta}). Rather, $T$ scales with $n$, which could be small even in the thermodynamic limit ($N\rightarrow\infty$).
% end revision

\begin{figure}[h!]
\begin{center}
\hspace{-0.7cm}
\includegraphics[scale=0.6]{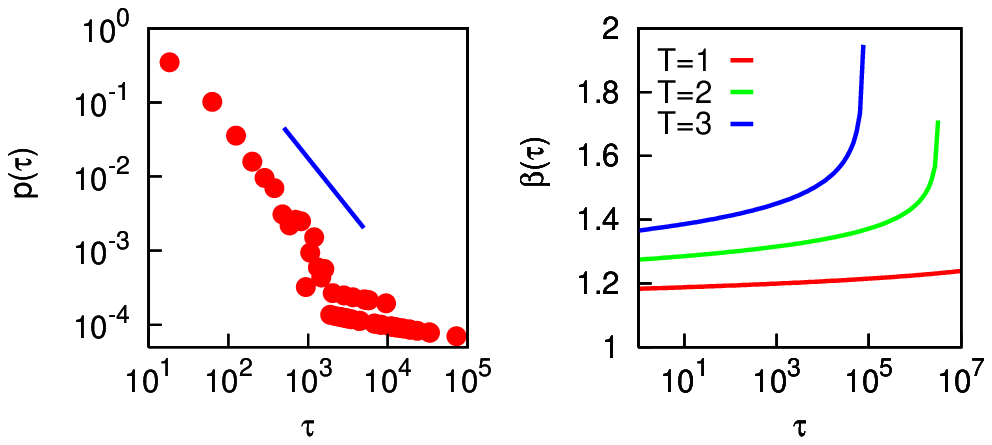}%{fig3.eps}
\end{center}
\caption{
%(colour online) 
Left panel: distribution of escape times $\tau$, as defined in the main text, for $\lambda=0.25$ and $T=2$, from MC simulations. Slope is for $\beta=1.35$. Other parameters as in Fig. \ref{fig_performance}.
Right panel: exponent $\beta$ of the quasi-power-law distribution $p(\tau)$ as given by Eq. (\ref{eq_beta}) for temperatures $T=1$, $2$ and $3$ (from bottom to top). 
}
\label{fig_aval_r}
\end{figure}

The left panel of Fig. \ref{fig_aval_r} shows the distribution of time intervals between events in which the overlap $m_{\mu}$ of at least one module $\mu$ changes sign. The power-law-like behaviour is apparent, and justifies talking about \textit{forgetting avalanches} -- since there are cascades of many forgetting events interspersed with long periods of metastability.
This is very similar to the behaviour observed in other nonequilibrium settings in which power-law statistics arise from the convolution of exponentials, such as demagnetization processes \cite{Hurtado} or Griffiths phases on networks \cite{Munoz}.

It is known from experimental psychology that forgetting in humans is indeed quite well described by power laws \cite{Wixted_1,Wixted_2,Sikstrom} -- although most experiments to date seem to refer to slightly longer timescales than we are interested in here.
The right panel of Fig. \ref{fig_aval_r} shows the value of the exponent $\beta(\tau)$ as a function of $\tau$. Although for low temperatures it is almost constant over many decades of $\tau$ -- approximating a pure power law -- for any finite $T$ there will always be a $\tau$ such that the denominator in the logarithm of Eq. (\ref{eq_beta}) approaches zero and $\beta$ diverges, signifying a truncation of the distribution.

% Revised:
Note that we have considered the information stored in a pattern to be lost once the system evolves to any other energy minimum. However, this new pattern will be highly correlated with the original one, and it might be reasonable to assume that the system has to escape from a large number of 
energy minima, $L$, before the information can be considered to have been entirely forgotten. The time for this is $\tau^{sum}=\sum_{i=0}^L \tau_i$, where $\tau_i$ are 
independently drawn from Eq (\ref{eq_Pt}). If $L$ is sufficiently large, 
the distribution of times $\tau^{sum}$ will tend to a L\'evy distribution \cite{Levy_distrib}. In practice, 
these different broad-tailed distributions [power-law, L\'evy, or as given by Eq. (\ref{eq_Pt})] are likely to be indistinguishable experimentally unless it is possible to observe over many orders of magnitude.
% end revision

%\subsection*{Clustered networks and spiking neurons}
\subsection*{Clustered networks}
\label{sec_clustered_nets}

Although we have illustrated how the mechanism of Cluster Reverberation works on a modular network, it is not actually necessary for the topology to have this characteristic -- only for the patterns to be in some way ``coarse-grained,'' as described, and that each region of the network encoding one bit have a small enough parameter $\lambda$, defined as the proportion of synapses to other regions. 
For instance, for the famous Watts-Strogatz \textit{small-world} model \cite{Watts} -- a ring of $N$ nodes, each initially connected to its $k$ nearest neighbours before a proportion $p$ of the edges are randomly rewired -- we have $\lambda\simeq p$ (which is not surprising considering the resemblance between this model and the modular network used above). More precisely, the expected modularity of a randomly imposed box of $n$ neurons is
$$
\lambda=p-\frac{n-1}{N-1}p+\frac{1-p}{n}\left(\frac{k}{4}-\frac{1}{2}\right),
$$
the second term on the right accounting for the edges rewired to the same box, and the third to the edges not rewired but sufficiently close to the border to connect with a different box.

Perhaps a more realistic model of clustered network would be a random network 
embedded in $d$-dimensional Euclidean space. For this we shall use the scheme laid out by Rozenfeld \textit{et al.} \cite{Rozenfeld}, which consists simply in allocating each node to a site on a $d$-torus and then, given a particular degree sequence, placing edges to the nearest nodes possible -- thereby attempting to minimize total edge length.
%\footnote{The authors also consider a cut-off distance, but we shall take this to be infinite here.} 
For a scale-free degree sequence (i.e., a set $\lbrace k_{i}\rbrace$ drawn from a degree distribution $p(k)\sim k^{-\gamma}$) according to some exponent $\gamma$, then, as shown in Analysis: Effective modularity of clustered networks,
%Materials and Methods,
%Appendix 1,
%\ref{appendix_benAv}, 
such a network has a modularity
\begin{equation}
\lambda\simeq\frac{1}{d(\gamma-2)-1}\left[d(\gamma-2)l^{-1}-l^{-d(\gamma-2)}\right],
\label{eq_lambda}
\end{equation}
where $l$ is the linear size of the boxes considered. It is interesting that even in this scenario, where the boxes of neurons which are to receive the same stimulus are chosen at random with no consideration for the underlying topology, these boxes need not have very many neurons for $\lambda$ to be quite low (as long as the degree distribution is not too heterogeneous).

\begin{figure}[h!]
\begin{center}
\includegraphics[scale=0.6]{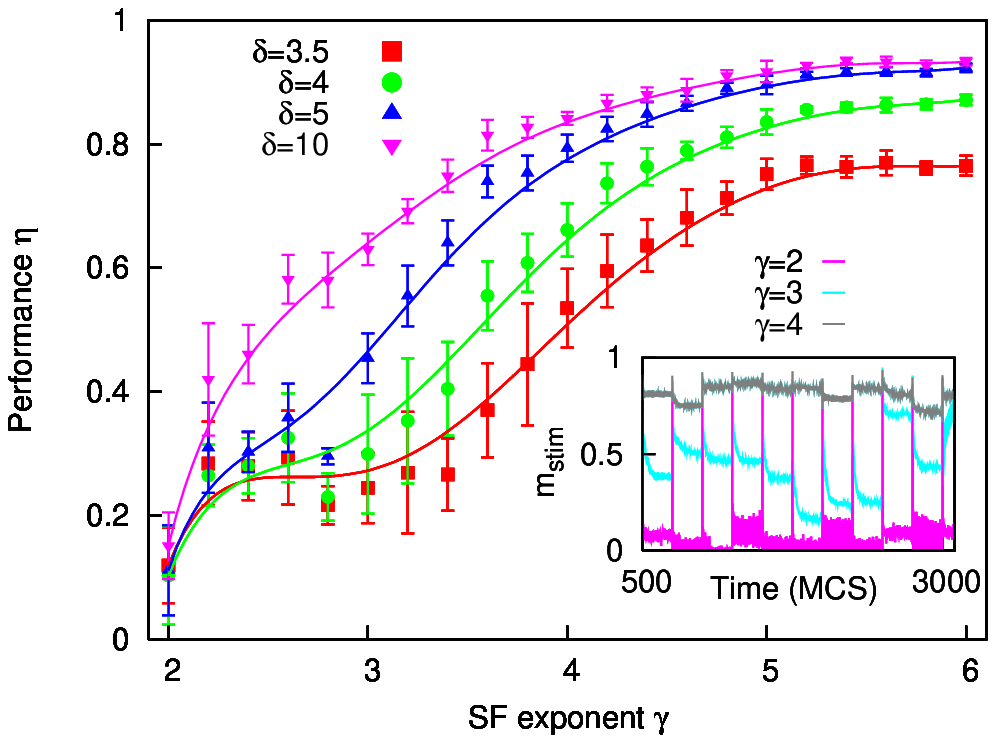}%{fig_6.eps}
\end{center}
\caption{Performance $\eta$ against exponent $\gamma$ for scale-free networks, embedded on a 2D lattice, with patterns of $M=16$ modules of $n=100$ neurons each, $\langle k\rangle=4$ and $N=1600$; patterns are shown with intensities $\delta=3.5$, $4$, $5$ and $10$, and $T=0.01$
(error bars represent standard deviations; lines -- splines -- are drawn as a guide to the eye). Inset: typical time series for $\gamma=2$, $3$, and $4$, with $\delta=5$.}
\label{fig_serie_SF}
\end{figure}

Carrying out the same repeated stimulation test as on the modular networks in Fig. \ref{fig_performance}, we find a similar behaviour for the scale-free embedded networks. This is shown in Fig. \ref{fig_serie_SF}, where for high enough intensity of stimuli $\delta$ and scale-free exponent $\gamma$, performance can, as in the modular case, be $\eta\simeq1$.
We should point out that for good performance on these networks we require more neurons for each bit of information than on modular networks with the same $\lambda$ (in Fig. \ref{fig_serie_SF} we use $n=100$, as opposed to $n=10$ in Fig. \ref{fig_performance}).
However, that we should be able to obtain good results for such diverse network topologies underlines that the mechanism of Cluster Reverberation is robust and not dependent on some very specific architecture.

\subsection*{Spiking neurons}

In the usual spirit of determining the minimal ingredients for a mechanism to function we have, up until now, used the simplest model neurons able to exhibit CR.
This approach makes for a good illustration of the main idea and allows for a
certain amount of analytical understanding of the underlying
phenomena. However, before CR can be considered as a plausible candidate for
helping to explain short-term memory, we must check that it is compatible with
more realistic neural models. For this we examine the behaviour of the popular
Integrate-and-Fire (IF) model neurons -- often referred to as {\it spiking
  neurons} -- in the same kind of setting as described
above
%  in Section \ref{sec_Cluster_Reverb}
for the simpler Amari-Hopfield neurons. In the IF
model, each neuron is characterized at time $t$ by a {\it membrane potential}
$V(t)$, described by the differential equation 
\begin{equation}
\tau_m\frac{dV(t)}{dt}=-V(t)+R_mI_{in}(t),
\end{equation}
% Joaquinito bit:
where $\tau_m$ and $R_{m}$ are, respectively, the membrane time constant
  and resistance, and $I_{in}(t)=I_{syn}(t)+I_{st}(t)+I_{ext}(t)$; the term $I_{syn}(t)=\sum_jI_{syn}^j(t)$ is the synaptic current
  generated by the arrival of Action Potentials (AP) from the neuron's presynaptic
neighbours, $I_{st}(t)$ is the current generated by the presentation of a
particular external stimulus to the network and $I_{ext}(t)=I_0+\sqrt{\tau_m}D\xi(t)$ is
an additional noisy external current. Here $I_0$ and $D$ are constants and
$\xi(t)$ is a Gaussian noise of mean $\langle\xi(t)\rangle=0$ and autocorrelation
$\langle\xi(t)\xi(t')\rangle=\delta(t-t')$. Each synaptic contribution to the
total synaptic current is modelled as  
$I_{syn}^j(t)=Ay_j(t)$, where $y_j(t)$ represents the fraction of
neurotransmitters in the synaptic cleft, which follows the dynamics \cite{tm97}
\begin{equation}
\frac{dy_j(t)}{dt}=-\frac{y_j(t)}{\tau_{in}}+U\delta(t-t_{sp}^j).
\end{equation}
Here, $t_{sp}^j$ is the time at which an AP arrives at synapse $j$, inducing the release of a fraction $U$ of neurotransmitters, and $\tau_{in}$ is
the typical time-scale for neurotransmitter inactivation.
% End Joaquinito
 Whenever $V$ surpasses a
given threshold $\theta$, the neuron fires an AP to all its postsynaptic
neighbours and $V$ is reset to zero, then undergoing a refractory
time $\tau_{ref}$ before again becoming susceptible to input. 
%(The IF model is described in greater detail in Appendix 2.)
%\ref{appendix_IF}.)
Because the parameters and variables of this model represent measurable physiological quantities, it is possible to use it to make quantitative -- albeit tentative -- predictions about the timescales on which CR might be expected to be effective in a real neural system.

\begin{figure}[h!]
\begin{center}
\includegraphics[scale=0.6]{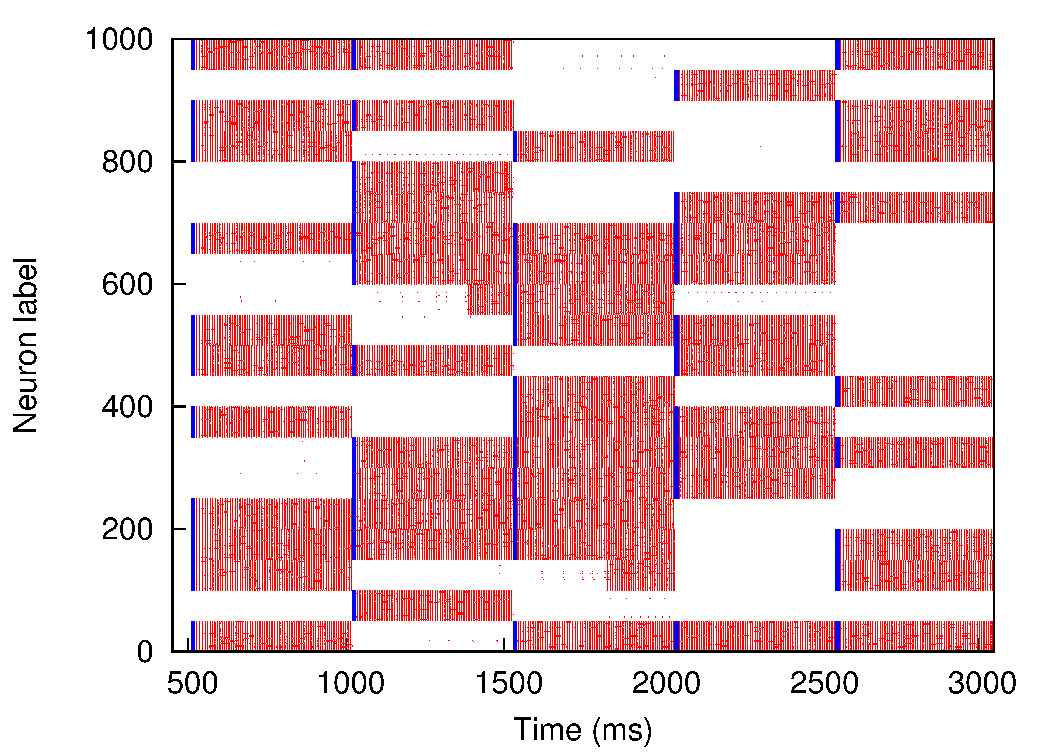}%{fig_if_ser.eps}
\end{center}
\caption{Raster plot, obtained from MC simulations, of a network of $1000$ integrate-and-fire (IF) neurons wired up (as described in the main text) in groups of $50$, with a rewiring probability $\lambda=0.02$. Every $500$ ms, a new pattern is shown for $10$ ms with an intensity $I_{stim}=500$ pA (plotted in blue). 
Parameters for the neurons are $A=42.5$ pA, $\theta=8$ mV, $\tau_{ref}=5$ ms, $\tau_{in}=3$ ms, $U=0.02$, $R_{m}=0.1$ $G\Omega$ and $\tau_m=10$ ms, which are all within the physiological range; and the external noisy current is modelled with $I_0=15$ pA and $D=10$ pA ms$^{-1/2}$.
}
\label{fig_raster}
\end{figure}

Figure \ref{fig_raster} is a raster plot of a modular network of IF neurons. The system performs a short-term memory task akin to the one previously described for the Amari-Hopfield neural network: the neurons belonging to clusters that correspond to ones in a random pattern are stimulated, for $10$ ms, with an intensity $I_{stim}$, while the the remaining neurons receive an opposite stimulus, $-I_{stim}$. We then allow the system to evolve for $500$ ms, before choosing a new random pattern and stimulating again. In such tests, the neurons in positively stimulated clusters usually begin to oscillate in synchrony, while the rest remain silent (save for occasional individual APs caused by noise). However, since this is a metastable state, with time active clusters can suddenly go mostly silent, or the neurons in silent clusters begin spontaneously to fire in synchrony. Thus, the information is gradually lost, as in the case with simpler neurons.

\begin{figure}[h!]
\begin{center}
\includegraphics[scale=0.6]{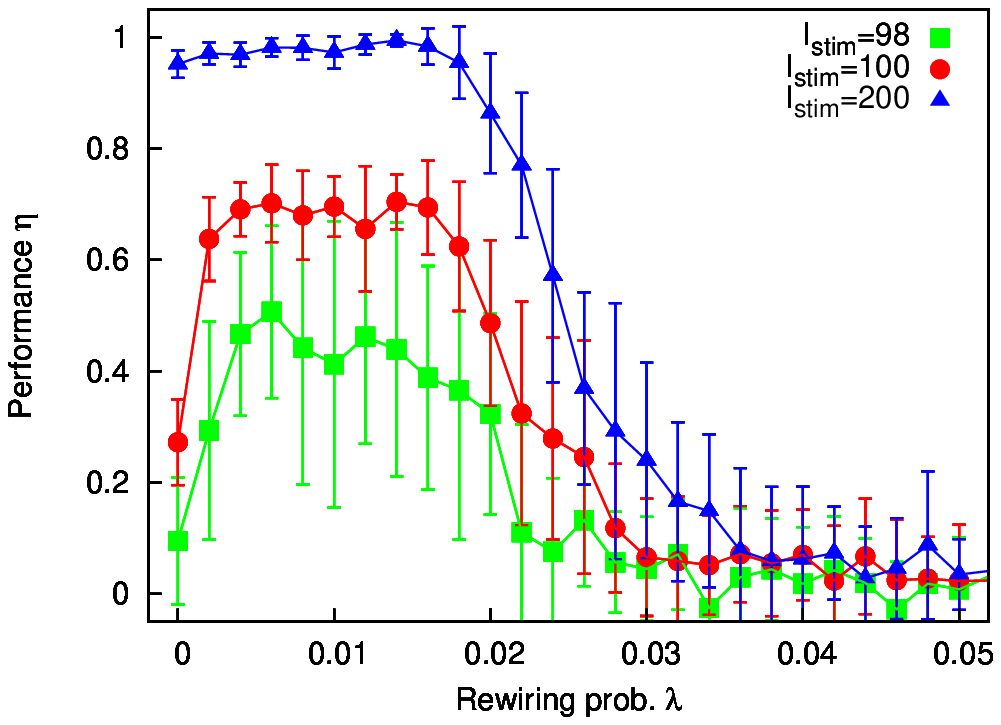}%{fig_if_ser.eps}
\end{center}
\caption{Performance $\eta$ against rewiring $\lambda$ for modular networks of IF neurons, as obtained from MC simulations. The network is periodically stimulated with a new random pattern for $10$ ms with an intensity $I_{stim}=98$ pA (green squares), $100$ pA (red circles) and $200$ pA (blue triangles) (error bars represent standard deviations; lines -- splines -- are drawn as a guide to the eye). The system evolves in the absence of stimuli for $5000$ ms and performance, $\eta$,
is computed according to Eq. (\ref{eq_eta_IF}).
%(defined in the main text) is measured over the final $100$ ms.
(An interval of $5$ seconds corresponds roughly to the timescale on which short-term memory operates in the brain.) Other parameters are as in Fig. \ref{fig_raster}.
}
\label{fig_serie_IF}
\end{figure}

To gauge how well the system is performing the task, we look at each cluster $\mu$ for the last $100$ ms before the next stimulus and assign a value $m_{\mu}=1$ to its mean activity if it is active, and $m_{\mu}=-1$ if it is silent. We then define the performance as:
\begin{equation}
\eta=\frac{1}{M}\sum_{\mu}m_{\mu}\xi_{\mu}.
\label{eq_eta_IF}
\end{equation}
In Fig. \ref{fig_serie_IF} we show the values of $\eta$ obtained in MC simulations against $\lambda$. Using different values of $I_{stim}$ we observe a similar behaviour to that of Fig. \ref{fig_performance}. In particular, for $I_{stim}\simeq 100$ pA, we have the interesting nonmonotonic behaviour in which performance benefits from a certain degree of rewiring. While, for the sake of illustration, in Fig. \ref{fig_raster} we only show the evolution of the system for $500$ ms after stimulation, in Fig. \ref{fig_serie_IF} we wait for five seconds.
% Revised:
Although the model is too simple, and the network too small, to make quantitative predictions about the brain, it is nevertheless promising that with physiologically realistic parameters we observe high performance ($\eta\simeq 1$) over several seconds, since this is the timescale on which short-term memory operates in humans.
\section*{Discussion}
\label{sec_discu}

%Cluster Reverberation is a means available to neural systems for robust short-term memory without synaptic learning. To the best of our knowledge, this is the first mechanism proposed which has these characteristics -- essential for, say, sensory memory or certain working-memory tasks.

Cluster Reverberation may be a means available to neural systems for performing certain short-term tasks, such as sensory memory or working memory. To the best of our knowledge, it is the first mechanism proposed to use network properties with no need of synaptic learning. All that is required is for the underlying network to be highly clustered or modular, and for small groups of neurons in some sense to store one bit of information, as opposed to a conventional view which assumes one bit per neuron. Considering the enormous number of neurons in the brain, and the fact that real neurons are possibly too noisy to store information individually anyway, these hypotheses do not seem far-fetched. The mechanism is furthermore consistent with what is known about the structure of biological neural networks, with experiments that have revealed power-law statistics of forgetting, and with recent observation of locally synchronized synaptic activity.

For the sake of illustration, we have focused here on the simplest model neurons that are able to exhibit the behaviour of interest. However, we have shown how the mechanism can also work with the slightly more sophisticated Integrate-and-Fire neurons, and there is no reason to believe that it would not also be viable with more realistic models, or even actual cells. Although CR comes about thanks to the high modularity of small groups of neurons, we have shown how robust it is to the details of the topology by carrying out simulations on clustered networks with no explicitly built-in modularity. And this setting suggests an interesting point. If an initially homogeneous (i.e., neither modular nor clustered) area of brain tissue were repeatedly stimulated with different patterns in the same way as we have done in our simulations, then synaptic plasticity mechanisms (LTP and LTD) might be expected to alter the network structure in such a way that synapses within each of the imposed modules would all tend to become strengthened, while inter-module synapses would vary their weights in accordance with the details of the patterns being shown \cite{Ole}. The result would be a modular structure conducive to efficient CR for arbitrary patterns, with simultaneous Hebbian learning in the inter-synapses of the actual patterns shown. In this way, the same network might be capable of both short-term and long-term memory, explaining, perhaps, why our brains can indeed store completely novel information but usually with a certain bias in favour of what we are expecting to perceive.

Although we have not gone into the question of neural coding, there would seem to be an intrinsic difference between \textit{semantic} storage of information -- used for long-term memory and probably useful for certain working-memory tasks that require the labelling of previously learned information --
and \textit{sensory} storage, for which some mechanism such as the one proposed here must store novel information immediately -- in a similar but more efficient way to how the retina retains the pigmentation left by an image it was recently exposed to.
If novel sensory information were held for long enough in metastable states, Hebbian learning (either in the same or other areas of the brain) could take place and the information be stored thereafter indefinitely. This might constitute the essence of concentrating so as to memorise a recent stimulus.

Finally, we should mention that CR could work in conjunction with other mechanisms, such as processes leading to cellular bistability, making these more robust to noise and augmenting their efficacy. Whether CR would work for biological neural systems could perhaps be put to the test by growing such modular networks {\it in vitro}, stimulating appropriately, and observing the duration of the metastable states \cite{Ole2,Shein}. {\it In vivo} recordings of neural activity during short-term memory tasks, together with a mapping of the underlying synaptic connections, might be used to ascertain whether the brain could indeed harness this mechanism. For this it must be borne in mind that the neurons forming a module need not find themselves close together in metric space, and that effective modularity might come about via stronger intra- than inter-connexions, instead of simply through a higher density of synapses within the clusters.
We hope that observations and experiments such as these will be carried out and eventually reveal something more about the basis of this puzzling emergent property of the brain's known as thought.

\section*{Analysis}
\subsection*{The effect of noise}

On a random network ($\lambda=1-M^{-1}$), the Amari-Hopfield model described in the main text has a second order phase transition with temperature, $T$, at $T_c=\omega \overline{k}$ \cite{Amit}. This can be seen by considering the mean-field equation for the overlap at the steady state, $m=\tanh(\omega \overline{k}m/T)$, where we have substituted $h_i=\omega \sum_j a_{ij} s_i \rightarrow \omega\overline{k} m$ in Eq. (\ref{eq_P}). For $T<T_c$, the paramagnetic solution $m=0$ becomes unstable, and ferromagnetic solutions $m\neq 0$ appear \cite{Fraiman}. This result also holds for the modular networks described in the main text. However, that the global overlap $m$ is different form zero does not mean that the short-term memory configurations we are interested in are stable. In fact, we know they are metastable for any $T>0$ (see Section {\it Energy and topology}), but we can set an upper bound on the temperature at which these states can be maintained even for a short time by considering again the mean-field equation for such a configuration. For a neuron in module $\mu$, $h_i\rightarrow \omega\overline{k}[(1-\lambda)m_\mu + \lambda m]$. For patterns with mean activity zero ($m=0$), states $m_\mu\neq 0$ will be unstable if $T>(1-\lambda)\omega\overline{k}\leq T_c$.

\begin{figure}[htb!]
\begin{center}
\includegraphics[scale=0.5]{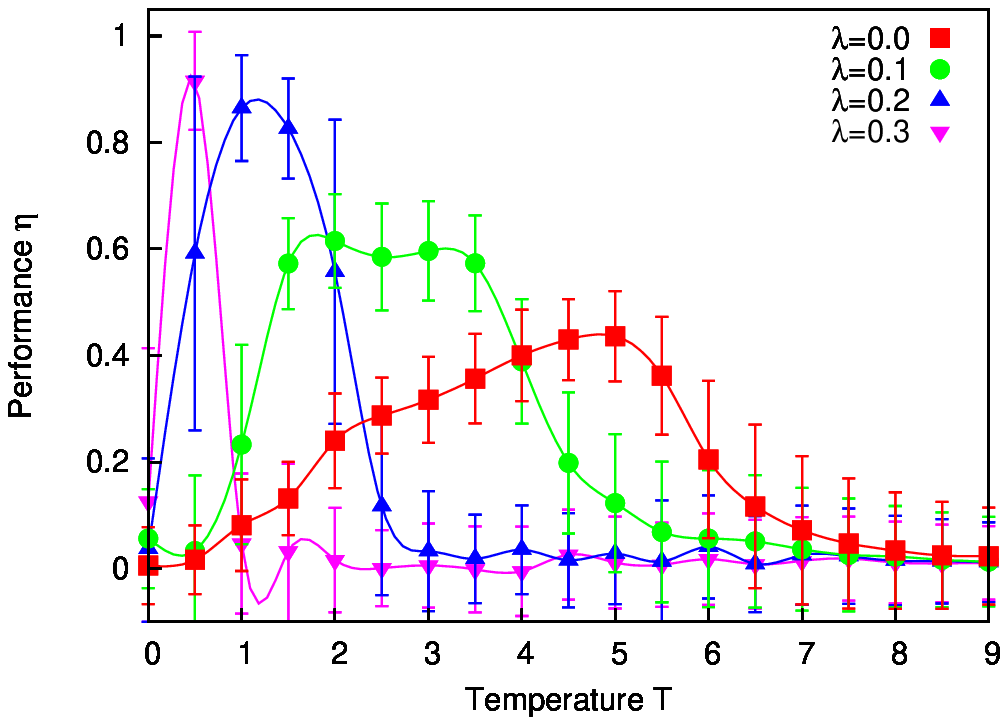}%{fig1.eps}
\end{center}
\caption{
Performance $\eta$ against $T$ for the Hopfield-Amari networks described in the main text, obtained from MC simulations, for values of the rewiring $\lambda=0.0$, $0.1$, $0.2$ and $0.3$, and stimulus $\delta=8.5$. All other parameters as in Fig. \ref{fig_performance}. (Error bars represent standard deviations; lines -- splines -- are drawn as a guide to the eye).
}
\label{fig_T}
\end{figure}

As we saw from Fig. \ref{fig_performance}, for stimuli $\delta\lesssim \omega\langle k\rangle$, the system does not always leave whichever meatastable state it is in to go perfectly to the pattern shown. A degree of ``structural noise'' ($\lambda>0$) can lead to a better response. In the same way, the dynamical noise set by $T$ can improve performance. Figure \ref{fig_T} shows how performance varies with $T$ for different values of $\lambda$. Due to the trade-off between sensitivity to stimuli and stability of the memory states, there is in general an optimum level of noise at which the system performs best. This dynamics can be interpreted as a kind of stochastic resonance, with the stimuli playing the part of the periodic forcing typically seen in such systems \cite{stoc_res}. Both the dynamic (annealed) noise, $T$, and the structural (quenched) noise, $\lambda$, serve to increase the sensitivity of the system to stimuli.

It is interesting to observe in Fig. \ref{fig_T} that whereas highly modular networks ($\lambda\simeq 0$) are most robust to $T$, for no values of parameters do they exhibit as good performance as the less modular networks when $T$ is relatively low.

%\section*{Appendix 1: Effective modularity of clustered networks}
%\section*{Materials and Methods}
\subsection*{Effective modularity of clustered networks}
\label{appendix_benAv}

\begin{figure}[h!]
\begin{center}
\includegraphics[scale=0.5]{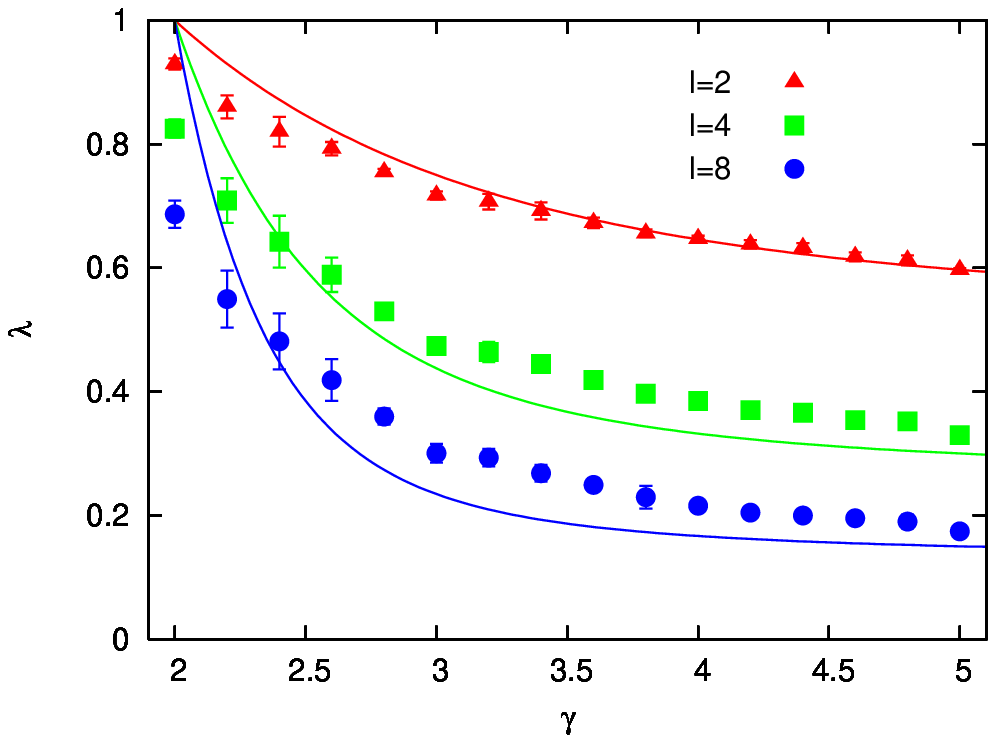}%{fig_7.eps}%{fig_5.eps}%{fig_mod_new_retake.eps}
\end{center}
\caption{
Proportion of outgoing edges, $\lambda$, from boxes of linear size $l$ against exponent $\gamma$ for scale-free networks embedded on $2D$ lattices. Lines from Eq. (\ref{eq_lambda})
and symbols (with error bars representing standard deviations) from simulations with $\langle k\rangle=4$ and $N=1600$.
}
\label{fig_mod_new4}
\end{figure}

We wish to estimate $\lambda$, the proportion of edges that cross the boundaries of a box of linear size $l$ placed randomly on a network embedded in $d$-dimensional space according to the scheme laid out in Ref. \cite{Rozenfeld}. The number of nodes within a radius $r$ is $n(r)=A_{d}r^{d}$, with $A_{d}$ a constant. We shall therefore assume a node with degree $k$ to have edges to all nodes up to a distance $r(k)=(k/A_{d})^{1/d}$, and none beyond (note that this is not necessarily always feasible in practice). To estimate $\lambda$, we shall first calculate the probability that a randomly chosen edge have length $x$. The chance that the edge belong to a node with degree $k$ is $\pi(k)\sim kp(k)$ (where $p(k)$ is the degree distribution). The proportion of edges that have length $x$ among those belonging to a node with degree $k$ is $\nu(x|k)=dA_{d}x^{d-1}/k$ if $A_{d}x^{d}<k$, and $0$ otherwise.
Considering, for example, scale-free networks (as in Ref. \cite{Rozenfeld}), so that the degree distribution is $p(k)\sim k^{-\gamma}$ in some interval $k\in [k_{0},k_{max}]$, and
integrating over $p(k)$, we have the distribution of lengths, 
$$%\begin{equation}
 P(x)=(Const.)\int_{\max(k_{0},Ax^{d})}^{k_{max}}\pi(k)\nu(k|x)dk=d(\gamma-2)x^{-[d(\gamma-2)+1]},
$$%\end{equation}
where we have assumed, for simplicity, that the network is sufficiently sparse that $\max(k_{0},Ax^{d})=Ax^{d}$, $\forall x\geq1$, and where we have normalised for the interval $1\leq x<\infty$; strictly, $x\leq(k_{max}/A)^{1/d}$, but we shall also ignore this effect. Next we need the probability that an edge of length $x$ fall between two compartments of linear size $l$. This depends on the geometry of the situation as well as dimensionality; however, a first approximation which is independent of such considerations is
$$%\begin{equation}
P_{out}(x)=\min\left(1,\frac{x}{l}\right).
$$%\end{equation}
We can now estimate the modularity
$\lambda$ as
$$%\begin{equation}
\lambda=\int_{1}^{\infty}P_{out}(x)P(x)dx=\frac{1}{d(\gamma-2)-1}\left[d(\gamma-2)l^{-1}-l^{-d(\gamma-2)}\right].
%\label{eq_lambda_app}
$$%\end{equation}
Figure \ref{fig_mod_new4} compares this expression with the value obtained numerically after averaging over many network realizations, and shows that it is fairly good -- considering the approximations used for its derivation.

\section*{Acknowledgements}
Many thanks to Jorge F. Mejias, Sebastiano de Franciscis, Miguel A. Mu\~noz, Sabine Hilfiker, Peter E. Latham, Ole Paulsen and Nick S. Jones for useful comments and suggestions. This work was supported by Junta de Andaluc\'{i}a projects FQM-01505 and P09-FQM4682, by the joint Spanish Research Ministry (MEC) and the European Budget for the Regional Development (FEDER) project FIS2009-08451, and by the Granada Research of Excellence Initiative on Bio-Health (GREIB) traslational project GREIB.PT\_2011\_19 of the Spanish Science and Innovation Ministry (MICINN) ``Campus of International Excellence.'' S.J. is grateful for financial support from the Oxford Centre for Integrative Systems Biology, and from the European Commission under the Marie Curie Intra-European Fellowship Programme PIEF-GA-2010-276454.

%This work was supported by Junta de Andaluc\'{i}a projects FQM-01505 and P09-FQM4682, and by Spanish MEC-FEDER project FIS2009-08451 and MICINN CEI-GREIB translational project GREIB.PT\_2011\_19. S.J. is grateful for financial support from the Oxford Centre for Integrative Systems Biology, and from the European Commission under the Marie Curie Intra-European Fellowship Programme PIEF-GA-2010-276454.

%\newpage

\end{document}